# Thermoelectric Signal Enhancement by Reconciling the Spin Seebeck and Anomalous Nernst Effects in Ferromagnet/Non-magnet Multilayers

Kyeong-Dong Lee[1], Dong-Jun Kim[1], Hae Yeon Lee[1], Seung-Hyun Kim[2], Jong-Hyun Lee[2], Kyung-Min Lee[2], Jong-Ryul Jeong[2], Ki-Suk Lee[3], Hyon-Seok Song[4,5], Jeong-Woo Sohn[4,5], Sung-Chul Shin[4,5], and Byong-Guk Park[1*]

[1]*Department of Materials Science and Engineering, KI for the Nanocentury, KAIST, Daejeon, 305-701, Korea*

[2]*Department of Materials Science and Engineering, Graduate School of Green Energy Technology, Chungnam National University, Daejeon, 305-764, Korea*

[3]*School of Mechanical and Advanced Materials Engineering, UNIST, Ulsan, 689-798, Korea*

[4]*Department of Physics and CNSM, KAIST, Daejeon, 305-701, Korea*

[5]*Department of Emerging Materials Science, DGIST, Daegu, 711-873, Korea*

*Email: bgpark@kaist.ac.kr

Abstract

The utilization of ferromagnetic (FM) materials in thermoelectric devices allows one to have a simpler structure and/or independent control of electric and thermal conductivities, which may further remove obstacles for this technology to be realized. The thermoelectricity in FM/non-magnet (NM) heterostructures using an optical heating source is studied as a function of NM materials and a number of multilayers. It is observed that the overall thermoelectric signal in those structures which is contributed by spin Seebeck effect and anomalous Nernst effect (ANE) is enhanced by a proper selection of NM materials with a spin Hall angle that matches to the sign of the ANE. Moreover, by an increase of the number of multilayer, the thermoelectric voltage is enlarged further and the device resistance is reduced, simultaneously. The experimental observation of the improvement of thermoelectric properties may pave the way for the realization of magnetic-(or spin-) based thermoelectric devices.

Thermoelectric (TE) effects, the conversion of thermal energy to electric signal, have gained increasing attention because of their potentials for harvesting electric energy from various sources including waste heat[1-5]. The realization of the TE effect as a practical power source requires developing thermoelectric materials with enhanced conversion efficiency which is normally expressed by thermoelectric figure of merit, $ZT = S^2\sigma T\kappa^{-1}$, where $S$ is the Seebeck coefficient, $\sigma$ is the electric conductivity, $\kappa$ is the thermal conductivity, and $T$ is absolute temperature[6]. Therefore, the TE efficiency will be improved by an increase of the electric conductivity and a reduction of the thermal conductivity of the materials. However, those two conductivities are interrelated so that their independent control is a great challenge. It has been reported that the thermal conductivity can be reduced by the interface control of nanostructure or by the introduction of the superlattice while keeping a similar electric conductivity[7-9]. However, further enhancement has yet to be done.

Recently, a new type of the thermoelectrics has been discovered in ferromagnet (FM)/non-magnet (NM) bilayer structures, where a thermal gradient generates spin current which is converted to a detectable voltage[10-13]. Because this TE effect is based on the spin current, it is called spin thermoelectric (STE) or spin caloritronics. In such system, at least two materials are necessary, so that the electrical and thermal conductivities are not limited to the Wiedemann-Franz law, but permitted to be independently modulated. One example is to use FM insulator as a spin current source[14,15] where the electric properties are only dependent on the NM material while most of the temperature gradient is imposed on the FM insulator. Even in the metallic systems, a proper combination of the FM and NM layer can also optimize the electrical and thermal conductivities. In addition to the controllability of conductivities, the STE has other advantages over conventional TE device such as simpler device structure, scaling capability and versatile applicability to various substrates[14]. Nonetheless, for practical realization of STE technique, an overall enhancement of the signal level is needed.

In this work, we present two approaches to enhance the STE effect in FM/NM structures, where the temperature gradient is vertically imposed in longitudinal configuration. The first one is to find a proper selection of the FM and NM materials to have a matching sign of the spin Seebeck effect (SSE) and anomalous Nernst effect (ANE) which contribute to the total STE voltage[13,16,17]. The second is to introduce a FM/NM multilayer structure where the injection of the thermally induced spin current can be multiplied. This demonstrates that the material engineering in STE devices would enhance the TE signal as well as modulate the device resistance simultaneously.

**Results**

**Thermoelectric voltage (*V*) induced by vertical temperature gradient.** The SSE is one of the TE in FM/NM bilayer structures, where thermally-induced spin current from FM layer is converted to charge current via inverse spin Hall effect (ISHE) in NM layer, as shown in Fig. 1(a). The magnitude of the SSE is $E_{SSE} \propto S_S \nabla T_{FN} \sim \theta_{SH}(\boldsymbol{J}_s \times \boldsymbol{\sigma})$, where $S_S$ is spin Seebeck coefficient that is determined by spin Hall angle ($\theta_{SH}$) of the NM layer and $\nabla T_{FN}$ is temperature gradient between FM and NM that induces spin current ($J_s$), $\boldsymbol{\sigma}$ is spin polarization vector[16-19]. In addition, there is another TE contribution in this configuration, anomalous Nernst effect (ANE) which is one of the properties of the FM materials when a thermal gradient is applied to a normal direction of the magnetization (Fig. 1(a)). This is expressed as $E_{ANE} \propto C_{ANE}(\boldsymbol{M}_s \times \nabla T_F)$, where $C_{ANE}$ is ANE coefficient of FM, $\boldsymbol{M}_s$ is magnetization vector, and $\nabla T_F$ is temperature gradient within the FM layer[16,17,20]. Therefore, in a given magnetization direction and temperature gradient, the total STE voltage is composed of the SSE and the ANE which can be either additive or subtractive depending on the relative sign of the $S_S$ (or $\theta_{SH}$) and the $C_{ANE}$.

We have measured the STE signal by utilizing an optically-induced heating method where the laser illumination in the central area of the devices generates vertical temperature gradient of the FM/NM bilayer samples. The magnetization of the FM layer is controlled by an in-plane magnetic field applied to the transverse direction to the voltage contact, as shown in Fig. 1(a), unless otherwise specified. Figure 1(b) shows the magnetic field dependence of the TE voltage for samples of CoFeB (CFB)/Ta. The TE voltage ($\Delta V$) changes its sign upon a reversal of the magnetization, which demonstrates the magnetic origin of the observed TE signal. Figure 1(c) presents the linear dependence of $\Delta V$ on the laser power which generates the thermal gradient in the sample. Moreover, the angular dependence of $\Delta V$ is well fitted to a cosine function (Figure 1(d)), which is consistent with the ANE and SSE geometry. These data confirm that the $\Delta V$ results from the magneto-thermoelectricity (ANE and/or SSE) induced by laser heating.

**Numerical calculation of temperature gradient.** To understand the thermoelectric effect in FM/NM bilayer quantitatively, we performed numerical calculation of temperature profile of the CFB/NM bilayers using COMSOL software and the material parameters as listed in Table I[21,22]. Heat transfer equation was calculated with heat source of Gaussian laser beam with width ($\sigma_L$) of 31.2 μm and power of 17 mW, which penetrates the materials with an absorption coefficient α. The effect of laser power and its spot size was described in detail in the Supplementary Fig. S1. We used the area of 400×400 μm$^2$ and the thickness of Pt, Ta, or CFB of 15 nm, SiO$_2$ of 100 nm, and Si of 649.9 μm. Top surface to ambient radiation is considered at room temperature of 293.15 K. To estimate $C_{ANE}$ and $S_S$, we assume that the effective temperature[12] $T^*$ is approximately the same as the temperature $\langle T \rangle = \int_t T(z)dz / t$ of the layer, which is averaged vertically in the center position of the laser beam, where $t$ (15 nm) is the thickness of FM or NM. Here, $\nabla T_{FN}$ is defined with the gradient of the average temperature of FM ($<T_F>$) and NM ($<T_N>$) by ($<T_F>-<T_N>$)$t^{-1}$, because $t$ is also the average distance between FM and NM. Note that the change in the sign of spin accumulation (σ) was considered when the stack order was reversed. Numerically calculated temperature profile of the FM/NM bilayers will be described later.

**Thermoelectric voltage by means of locally-illuminated optical heating source.** Local optical induction of thermoelectric voltage reveals the dependence on the Hall-bar width of the CFB/Pt (a) and CFB/Ta (b) samples after eliminating the offset voltage, as shown in Fig. 2. The narrowest sample ($w = 0.2$ mm), i.e., with the highest resistance is observed to have a larger signal than samples with wider widths. Similarly, highly resistive CFB/Ta samples show a larger signal than the CFB/Pt samples. To clarify the resistance ($R$) dependence, we plotted the relationship between $\Delta V$ and $R$ in Figs. 2(c)-(e). Both the CFB/Pt and CFB/Ta samples show a linear dependence indicating that the combined signals of the anomalous Nernst effect (ANE) and the spin Seebeck effect (SSE) are linearly scaled with $R$. Since changing the width of the Hall-bar does not alter the relative resistivity of FM/NM bilayer, the result stems not from the shunting effect between FM and NM but from the local heating. A fixed size of the laser beam kept the excitation width ($d$) less than the total width ($w$) of the stripe. Therefore, the thermally-generated voltage ($V_S$) is laterally shunted by the resistance $R_w$ of the un-excited areal width ($w - d$) of the stripe, thereby resulting in the measured voltage $\Delta V = (dw^{-1})V_S$, as depicted in the circuit model of Fig. 2(e) [16]. Since SSE (ANE) is generated in NM (FM), the $V_S$ is expressed by $V_S = (I_{SSE} + I_{ANE}) \times R_S$, where $I_{SSE} \equiv V_{SSE}R_N^{-1}$, $I_{ANE} \equiv V_{ANE}R_F^{-1}$, and $R_S = R_F R_N (R_F + R_N)^{-1}$. Note that $R_F$ and $R_N$ are resistances for the excited width $d$ of FM and NM respectively, and the $I_{SSE}$ and $I_{ANE}$ are the equivalent current in the circuit model[23]. Since $d < w$, the $I_{SSE}$ and $I_{ANE}$ in our sample configuration are thought to be identical with the variation of the stripe width. Considering that the total electrical resistance $R = (dw^{-1})R_S$, the measured voltages ($V = (I_{SSE} + I_{ANE}) \times R$) could be simply proportional to $R$. Hence, in contrast to a case in which the entire sample area is excited[23], the thermoelectric signals generated by local heating could be normalized by total electrical resistance to investigate the ANE and SSE contribution. Namely,

$$\Delta V R^{-1} = E_{ANE}\, dR_F^{-1} + E_{SSE}\, dR_N^{-1}. \qquad (1)$$

Considering that $E_{ANE} = C_{ANE} M_s \nabla T_F$, and $E_{SSE} = -S_S \nabla T_{FN}$, where $C_{ANE}$ is the anomalous Nernst coefficient, and $S_S$ is the phenomenological spin Seebeck parameter[16,24], Eq. (1) can be expressed by

$$\Delta V R^{-1} = (C_{ANE} M_s \nabla T_F)(dR_F^{-1}) - (S_S \nabla T_{FN})(dR_N^{-1}). \qquad (2)$$

Here, $\nabla T_{FN}$ is the effective temperature gradient between FM and NM. The equation (2) describes the essential parameters to understand the measured $\Delta V R^{-1}$.

**Thermoelectric effect of heterostructures with different NM materials and inversing stack order.** In order to explore the relative contribution of the ANE and the SSE, we examined the TE effects of CFB and Co samples with different NM materials of Ta and Pt, as such materials are generally known to have opposite spin Hall angles[25-28]. This is confirmed by a ferromagnetic resonance spin pumping experiment with the samples grown in nearly identical conditions[28]. Figure 3 shows the TE voltages of the CFB/NM (a), NM/CFB (b), and Co/NM (c) samples, which are normalized by the total resistance. We found two particular points in the results. The first one is that the STE voltages are the same sign independent of the NM materials even though the SSE voltage is expected to be reversed with an NM material of an opposite $\theta_{SH}$ or $S_S$. As this trend is independent from its stack order for each NM material, as shown in Figs. 3(a) and (b), the inversion of the stack order may result in the same sign of SSE. This could be understood that the inversion of the stack order induces changes in the direction of $\boldsymbol{J}_s$ and in the sign of spin accumulation ($\boldsymbol{\sigma}$) simultaneously. Secondly, the FM/Pt samples show a larger signal than the FM/Ta samples irrespective of FM materials. Considering the similar thermal and optical properties of Pt and Ta as well as the same magnetic materials used, the ANE contribution from $\nabla T_F$ is expected to be similar for the samples with different NM materials.

Figures 3(d)-(f) shows the numerically calculated temperature profile of the CFB/NM and Co/NM bilayers. It is exhibited that the overall temperature profile of the bilayer in the same configuration is not significantly affected by changing the NM materials from Pt to Ta. The temperature gradient of CFB ($\nabla T_F$) and the effective temperature gradient between CFB and NM ($\nabla T_{FN}$) are summarized in Table II, together with the experimentally measured thermal voltages and resistances, since these are also used for the estimation of $C_{ANE}$ and $S_S$, as described in Eq. (2). In the calculation of FM (NM) resistance of $R_F$ ($R_N$), the resistivity (193.1 μΩ·cm) of single FM (*w*: 0.4 mm, $R_{CFB}$: 579.3 Ω) is thought to be the same as that of FM in bilayers. The resistivity of Pt and Ta is calculated to be 15.42 (16.16) and 162.6 (169.5) μΩ·cm for top NM (bottom NM) structure, respectively. The relatively high resistivity of Ta implies that Ta is in the β-phase. We would like to note that the top- and bottom-NM structures show similar resistance.

Intriguingly, the nearly same temperature gradient of top FM layer ($\nabla T_F$) in Fig. 3(e) implies the essential role of SSE for the signal difference of Fig. 3(b) due to the nearly same ANE contribution of NM/CFB structures. Thus, the dissimilar TE voltage between samples with Pt and Ta is possibly explained by the different SSE contribution. The similar phenomenon is also observed in the samples with a Co FM layer, as shown in Fig. 3(c). This implies that SSE can be additive (subtrative) to the ANE depending on the relative sign of the contributions.

**Spin Seebeck and anomalous Nernst coefficient of CoFeB/NM bilayer structures.** While the asymmetric $\nabla T_F$ and $\nabla T_{FN}$ is calculated by laser heating, we consider the material-dependent $C_{ANE}$ and $S_S$ as unchanging parameters with the stack order variation, but changing parameters with the different NM material, as described in Fig. 3. By using Table II and Eq.

(2) for the top and bottom Pt and Ta structures, we solve $C_{ANE, CFB(Pt, Ta)}$ and $S_{S, Pt, Ta}$ for Pt and Ta cases, respectively. The phenomenological spin Seebeck coefficients are obtained as $S_{S, Ta(CFB)}$ = -7.52×10$^{-7}$ V/K and $S_{S, Pt(CFB)}$ = 2.75×10$^{-7}$ V/K, which are comparable to those in literatures[15,16], $S_{S, Pt(YIG)}$ = 5.9×10$^{-8}$ V/K and 1×10$^{-7}$ V/K. The different signs of the $S_{S, Ta(CFB)}$ and $S_{S, Pt(CFB)}$ could be explained by the opposite sign of spin Hall angle, which determines the sign of $V_{ISHE}$. The $S_{S, Ta(CFB)}$ might be overestimated due to the unknown thermal conductivity of Ta ($\kappa_{Ta}$) in the samples. The $\kappa_{Ta}$ value used in the numerical calculation is a typical value of α-Ta. The measured lower electrical conductivity of Ta than bulk value implies that the actual $\kappa_{Ta}$ could be smaller according to the Wiedemann-Franz law.

The anomalous Nernst coefficient was found to be $C_{ANE, CFB(Pt)}$ = 3.28×10$^{-6}$ V/KT and $C_{ANE, CFB(Ta)}$ = 4.45×10$^{-6}$ V/KT, which are also comparable to the reported ones[16,20,29], $C_{ANE, Co2FeAl}$ = 9.5×10$^{-8}$ V/KT, $C_{ANE, FePt}$ = 5.6×10$^{-7}$ V/KT, and $C_{ANE, MnGa}$ = -7.6×10$^{-7}$ V/KT. Our observation implies that the $C_{ANE}$ is affected by NM material, which might depend on the approximate temperature calculation with bulk thermal properties or magnetic proximity effect. The interference between ANE and SSE might be another possible consideration. Here, note that the sign of ANE of CFB is indeed same as that of Pt, but opposite to that of Ta.

**Enhancement of thermoelectric effect in an optimized ferromagnet/non-magnet multilayer microstructure.** Next, we try to find an optimum thickness of the NM materials which maximizes the SSE. Figure 4(a) shows the TE voltage ($\Delta V$) of CFB(8nm)/Pt samples as a function of Pt thickness ($t_{Pt}$), presenting a peak of the $\Delta V$ at 2~3 nm of the $t_{Pt}$ and a reduction for smaller or larger $t_{Pt}$. The decrease of the $\Delta V$ for larger $t_{Pt}$ can be attributed to the reduction of the sample resistance with $t_{Pt}$ (Fig. 4(b)) since the $\Delta V$ is proportional to the resistance. The resistance effect is simply removed by the normalization of the $\Delta V$ as shown in Fig. 4(c). The resultant TE effect of $\Delta VR^{-1}$ increases initially, which then saturates at larger $t_{Pt}$. This is attributed to the thickness dependence of the ISHE which is proportional to the

NM layer thickness up to its spin diffusion length of ~ 3 nm, where the signal starts to be saturated, and then becomes constant for a larger thickness[23,26,30]. The value of the spin diffusion length is consistent with the result of our ferromagnetic resonance experiment and those of the other groups[28,30]. These determine an optimum $t_{Pt}$ where the $\Delta V$ is maximum or $\Delta V R^{-1}$ is saturated. Note that the ANE effect is assumed to be identical for all samples because of the same CFB thickness.

The STE has been mostly investigated in FM/NM bilayer structures, where in the vertical temperature gradient, the ANE and SSE can be additively combined depending on the sign of the $\theta_{SH}$ and ANE coefficient as demonstrated above. A larger TE signal could be obtained in the bilayer sample by an introduction of a FM material of a larger ANE coefficient and/or an NM material of a larger $\theta_{SH}$ ($S_S$). However, once the materials are selected, there is no way to improve the TE signal. Here, we present a further enhancement of the TE effect by the formation of the FM/NM multilayer. The multilayered samples were fabricated by a repetition of the CFB(8nm)/Pt(3nm) bilayer up to 10 time and their TE voltage ($\Delta V$) and resistance was measured. It can be simply expected that the formation of the $n$ multilayer make $n$ time of the TE effect ($\Delta V R^{-1}$) and $n^{-1}$ time of sample resistance ($R$) of the bilayer. Thus, the resultant TE voltage ($\Delta V$) will be constant irrespective of $n$. However, this is not the case. Interestingly, Fig. 5 shows that as the number of bilayer ($n$) is increased, the $\Delta V$ is continuously enhanced while their resistances are reduced with a relation of $n^{-1}$. As a result, the $\Delta V R^{-1}$ is enlarged with a proportion of $n^{\alpha}$ ($\alpha > 1$).

In Fig. 5(b), reduction of lateral shunting by reducing the stripe width from 100 μm to 10 μm increases the thermal voltage ($\Delta V$) by 10 time, when the power of laser beam is same, and its spot size is less than the stripe width, as described in Fig. 2. This means that lateral microstructuring can provide the nearly same TE effect ($\Delta V R^{-1}$) unless the vertical configuration of multilayer structure changes, and the signal mostly stems from the locally-generated vertical thermal pumping.

Here, by the multilayer stacking of 10 and by the stripe-width reduction of 1/10, the power factor ($\sigma S^2$) was enhanced by a factor of ~300, and $ZT$ was roughly estimated to be $1.73\times10^{-3}$ (see Supplementary information for detail). We think that there is a room to be further enhanced by lateral piling method.

While the exact origin of the enhancement of multilayer heterostructure remains yet to be understood, this can be inferred to the fact that the thermally induced spin current can be injected into both upper and lower NM material in the multilayer system, which may increase the STE effect more than simple multiplication as shown in Fig. 5(a). Moreover, a reduced thermal conductivity by the formation of the multiple interfaces can possibly contribute to the STE enhancement of the multilayer[31]. Importantly, optical absorption of metal layer enhances its thermal gradient induced by laser illumination. Figure 6(a) shows the relative temperature variation of [CoFeB(8 nm)/Pt(3 nm)]$_{10}$ multilayer sample when $P = 27$ mW and $d = 5$ μm of the experimental condition. Its average temperature gradient, as defined previously, is shown in Fig. 6(b) as a stack unit. Interestingly, this figure reveals that the exponential attenuation of optical source accelerates the rate of temperature gradient as the optical beam penetrates multilayer sample up to the number of repeat ($n$) ~ 4. This explains the rapid increase of $\Delta V$, when $n$ is less than 3~4. The $\nabla T$ dependence on the various optical absorption coefficients was given in the Supplementary information. One of reasons for the additional increase of $\Delta V$, even when $n > 6$ in Fig. 5(b), could be interpreted by the injection as well as extraction of spin current of NM layer. Overall, this means that ANE as well as SSE could be further enhanced in case of multilayer with optical heating.

**Discussion**

The exact ANE signal of the CFB cannot be separately quantified from the total measured TE voltages due to the same symmetry with SSE, as the **σ** is determined by the $M_s$, and $J_s$ is induced by the temperature difference between the FM and NM layer[16,17]. As a result, one could produce an analysis by interpreting the voltage changes in various samples as the result of changes in the ANE signal of the FM layers driven by the proximity to large spin orbit coupling materials. Therefore, $C_{ANE}$ and $S_S$ here might include magnetic proximity effect[32]. If the proximity effect existed with 1 nm of Pt[16,33], then approximately 7% (~ 1 nm/$t$) would be additionally gained to $C_{ANE, CFB(Pt)}$. Since $V/R$ of Ta/CFB corresponds to nearly 50% $V/R$ of Pt/CFB as shown in Fig. 3(b), the large signal difference reasonably indicates that in addition to ANE including proximity effect, ISHE could contribute to the signal difference between Pt/CFB and Ta/CFB, as the longitudinal SSE is clearly evidenced in ferromgnetic insulator with Pt electrode[17]. As addressed in the recent debate on the transverse SSE in metallic ferromagnets of permalloy and Ni[13,34-36], future studies like the angular dependence of STE voltages under well-controlled temperature gradient with various substrates as well as the interface quality treatment between FM and NM could shed light on complete understanding of longitudinal SSE in metallic FM/NM.

Neverthless, it is particularly interesting that the Pt (Ta) with a positive (negative) $S_S$ (or $\theta_{SH}$) can make the SSE be additive (subtractive) to the ANE of the CFB, which results in a larger thermoelectric voltage in Pt/CFB samples. As we assume the same $C_{ANE}$ and $S_S$ with the stack order variation, the estimated $C_{ANE}$ and $S_S$ represent the average value of the stack order variation even if the assumption is not strictly satisfied. We believe that our interpretation could be useful when $C_{ANE}$ and $S_S$ need to be determined in the coexistence of ANE and SSE by local heating method.

We would like to mention that contrary to the conventional Seebeck effect, the STE voltage ($\Delta V$) here is generated perpendicular to heat flow ($\nabla T_F$) direction, which allows more controllability. In addition, anomalous Nernst and spin Seebeck effect utilizes the magnetization in FM/NM multilayers, of which switching field could be widely tuned by the selection of materials and their thicknesses.

In conclusion, we have investigated the TE effect in FM/NM bilayer with NM materials with different $\theta_{SH}$ by means of an optical heating system. The TE effect of the samples with Pt is larger than those with Ta, which demonstrates that the TE effect contributed by the SSE and ANE can be improved by a selection of materials with a matching sign of the $S_S$ ($\theta_{SH}$) and the ANE coefficient. We also studied the [FM/NM]$_n$ multilayer samples and demonstrated an improvement of the TE voltage and the modulation of the resistance by formation of multilayer. This result illustrates that the formation of the multilayer allows one to have a large TE voltage and to be able to modulate resistance simultaneously, which is of great importance for practical application.

**Methods**

**Device fabrication.** Samples of FM/NM bilayer or [FM/NM]$_n$ multilayer structures were deposited on a thermally oxidized Si substrate by DC magnetron sputtering with a base pressure of $5\times10^{-8}$ Torr and a working pressure of 3 mTorr. Here, the FM materials are Co$_{32}$Fe$_{48}$B$_{20}$ (CFB) and Co, and the NM materials are Pt and Ta. Bar-shaped structures with widths of 0.01~0.5 mm and a length of 1.0~1.4 mm were defined using a photolithography and Ar ion milling or a shadow metal mask.

**Measurements.** A schematic of the experimental system used here is illustrated in Fig. 1(a). A 532-nm continuous or 800-nm pulsed laser with a power of 2~90 mW was used. A pulsed laser of 30 fs showed nearly identical magneto-thermoelectric signal to a continuous wave, but provided superior power stability and a better signal to noise ratio. The laser spot size was tuned to 5~300 μm with an objective lens depending on the device width, and the laser power was calibrated at the sample position. During the thermoelectric voltage measurement by a nanovoltmeter (Keithley 2182A), magnetization curve was simultaneously monitored by the MOKE using a lock-in amplifier with PEM modulation. The magnetic field was applied along the $x$ axis ($\varphi=0°$) and the laser spot was positioned at the center of the sample during each measurement, unless otherwise specified.

## Acknowledgements

This research was supported by Basic Science Research Program through the National Research Foundation of Korea (NRF) funded by the Ministry of Science, ICT & Future Planning (NRF-2012R1A1A1041590, 2013R1A2A2A01067144, 2013R1A4A1069528) and Education, Science and Technology (NRF-2013R1A1A2011103). It was also supported by the KAIST High Risk High Return Project (HRHRP). We would like to thank Dr. Y. M. Kim for critical reading.

## Author contributions

B.-G.P. and K.-D.L. conceived the experiments. D.-J.K. prepared the samples. K.-D.L. and D.-J.K. carried out experiments. S.-H.K., J.-H.L., K.-M.L. and J.-R.J. did simulations. H.-S.S, J.-W.S. and S.-C.S. contributed to the measurement setups. K.-S.L., B.-G.P., K.-D.L. and J.-R.J. analyzed the results. B.-G.P. and K.-D.L. wrote the manuscript with help of H.Y.L., K.-S.L. and J.-R.J..

## Additional information

Supplementary information accompanies this paper.

Competing financial interests: The authors declare no competing financial interests.

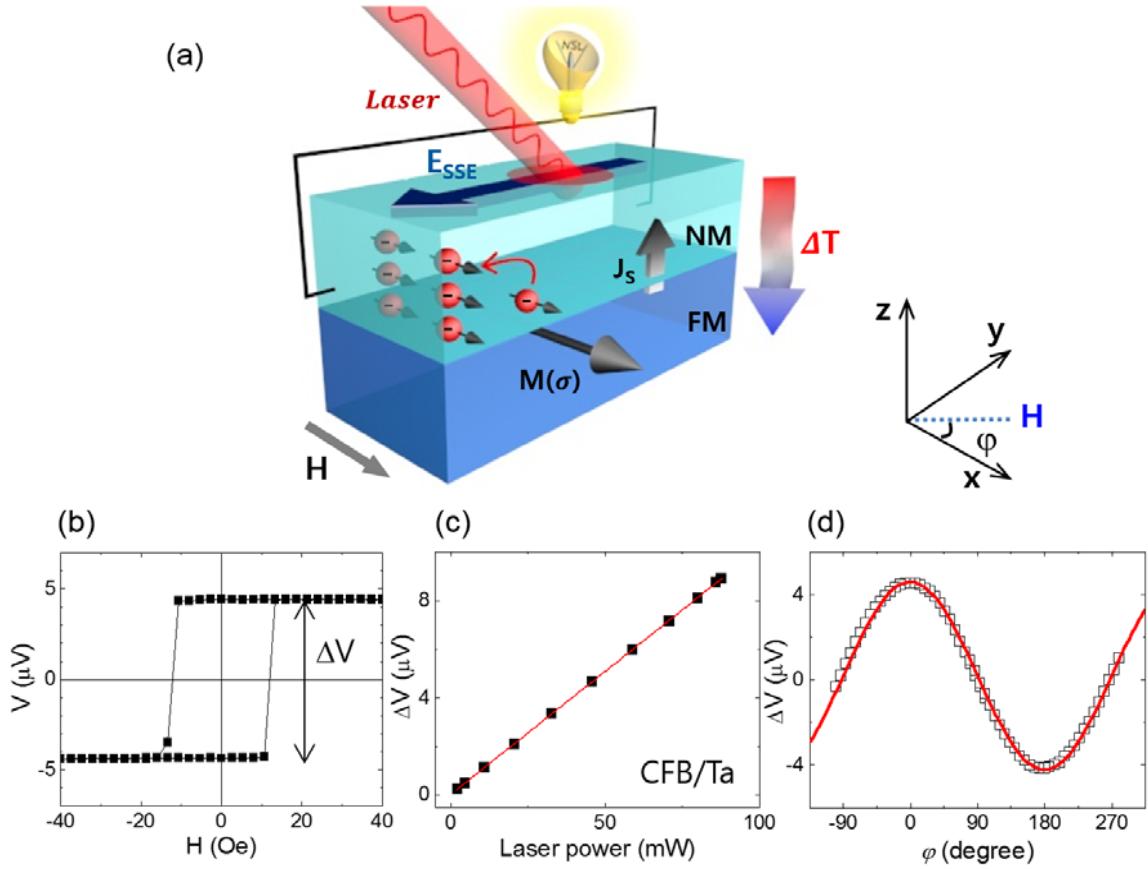

**Figure 1 | Magneto-thermoelectric effect in ferromagnet (FM)/non-magnet (NM) bilayer by illumination of laser.** (a) Schematic experimental configuration. The arrow in FM layer indicates the magnetization direction, and those in NM are spin orientation of the thermally injected electrons. (b) Thermoelectric voltage $V$ and magneto-thermoelectric voltage $\Delta V$ of the CFB/Ta sample of 0.4 mm width vs in-plane magnetic field. (c) Dependence of the $\Delta V$ on laser power intensity. The solid line is a linear fit. (d) The angular dependence of $\Delta V$ when applying $H$ with an in-plane angle of $\varphi$. The solid curve is a fit to cosine function.

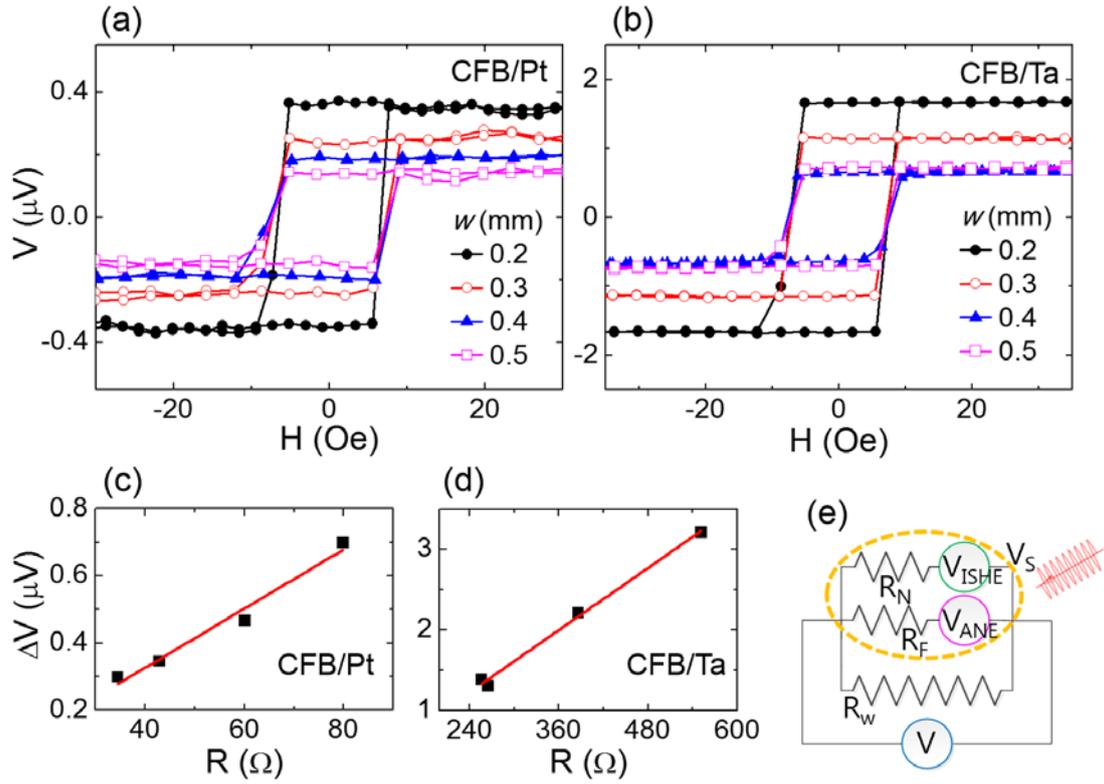

**Figure 2 | Locally laser-induced hybrid voltage (*V*) generation of SSE and ANE of CFB/NM with different hall-bar widths.** *V* of CFB/Pt (a) and CFB/Ta (b) at a laser power of 17 mW. Magneto-thermoelectric voltage ($\Delta V$) of CFB/Pt (c) and CFB/Ta (d) as a function of the Hall-bar resistance *R*. (e) Circuit model with FM resistance ($R_F$) and NM resistance ($R_N$). The un-excited lateral resistance is represented by $R_w$.

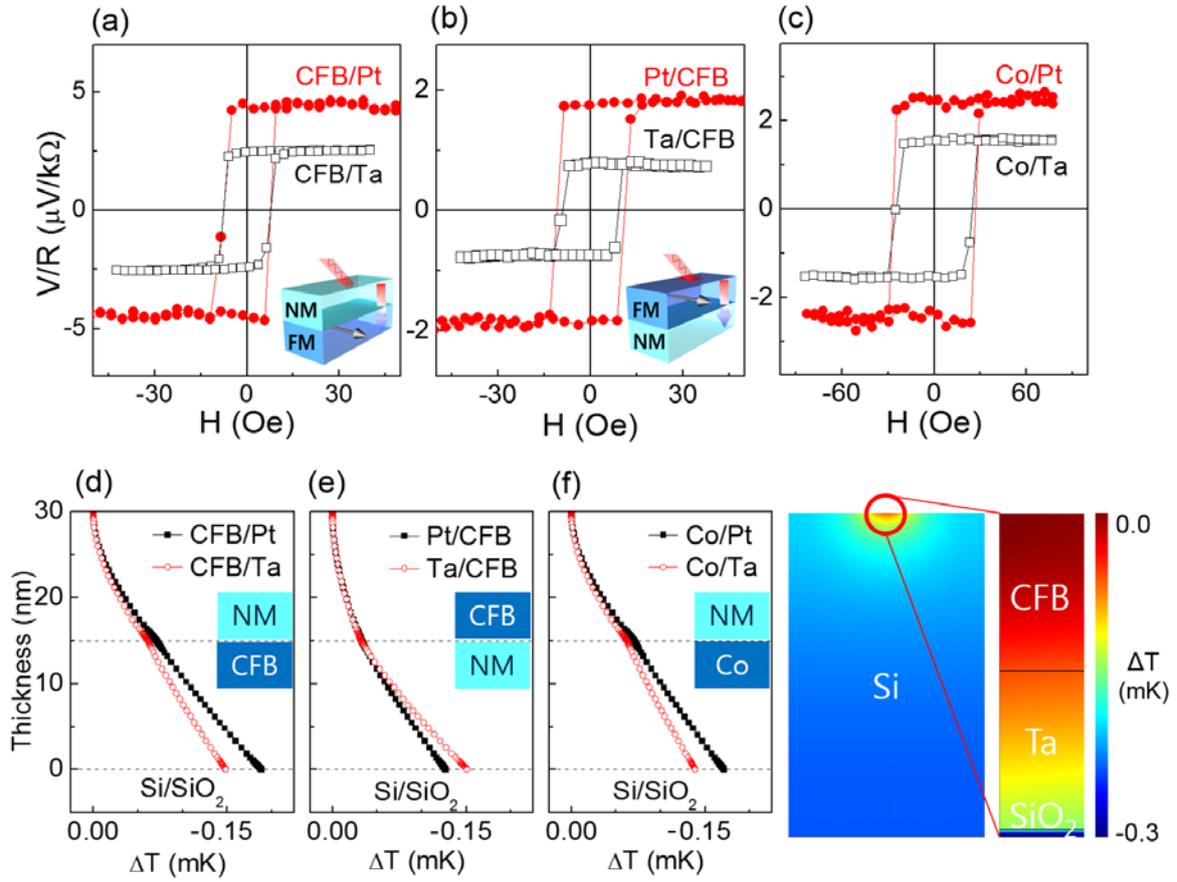

**Figure 3 | Normalized thermoelectric voltage as a function of magnetic field.** (a) CFB/NM, (b) NM/CFB, and (c) Co/NM samples, where NM is Pt or Ta. Numerically calculated temperature profiles of (d) CFB/NM, (e) NM/CFB, and (f) Co/NM samples.

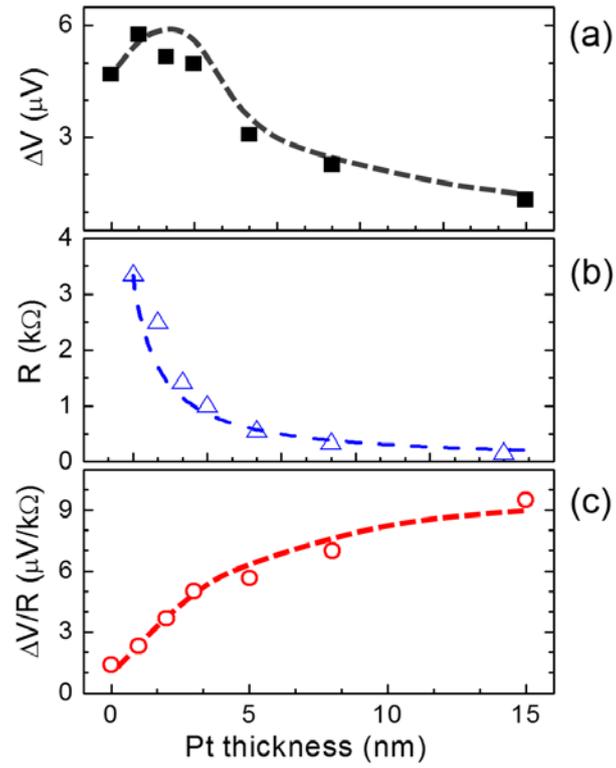

**Figure 4 | Pt thickness dependence on thermoelectric voltage ($\Delta V$).** (a) $\Delta V$, (b) resistance $R$, and (c) normalized thermoelectric voltage $\Delta V R^{-1}$, as a function of Pt thickness for the CFB/Pt samples of 0.1 mm width.

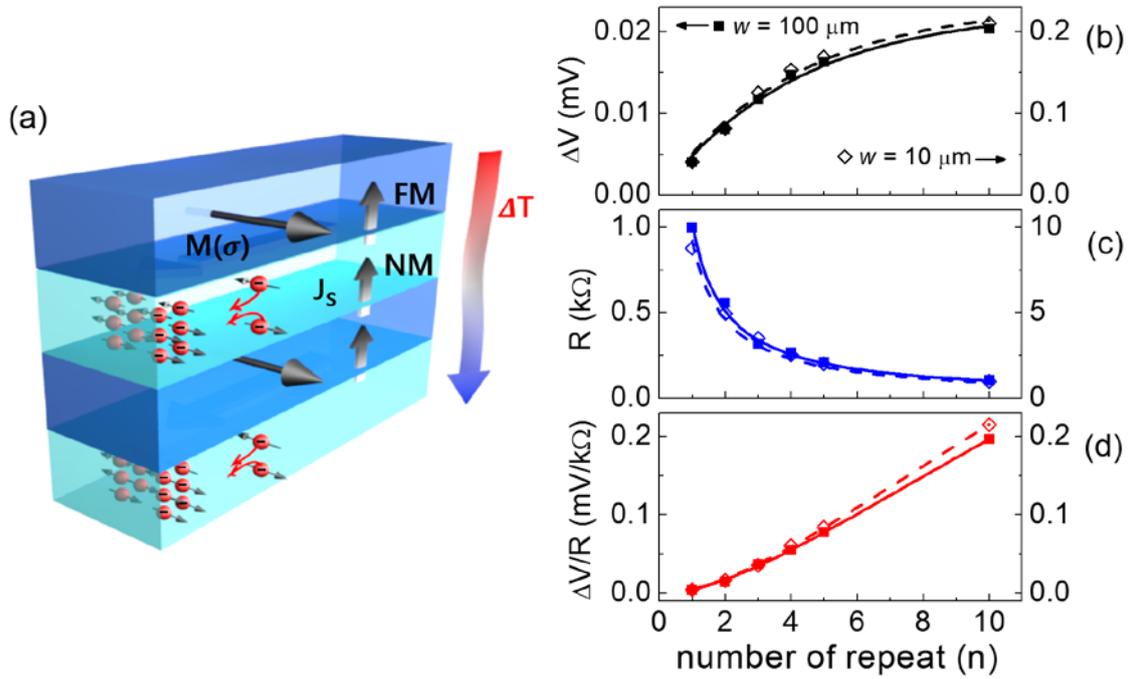

**Figure 5 | Enhancement of thermoelectric voltage by formation of [CFB/Pt]$_n$ multilayers.**
(a) Schematic description of the injection as well as extraction of spin current of NM layer, which is one of possible origins of the TE signal enlargement. (b) Thermoelectric voltage ($\Delta V$), (c) resistance ($R$), and (d) normalized thermoelectric voltage ($\Delta V R^{-1}$), as a function of the number of multilayer ($n$) for CFB/Pt samples of (left *y* axis) 0.1 mm and (right *y* axis) 0.01 mm width.

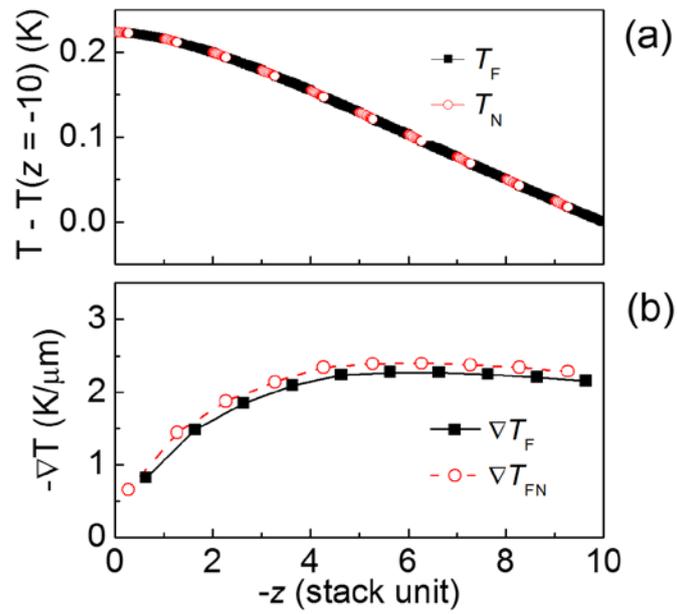

**Figure 6 | Numerically calculated temperature profile of [CoFeB(8 nm)/Pt(3 nm)]$_{10}$ multilayer when $d$ = 5 μm.** (a) Relative temperature variation. (b) Temperature gradient as a stack unit. The average temperature gradient for each stack is shown. Laser beam is illuminated from left where $z = 0$.

**Table I** | Material parameters for numerical calculation of temperature gradient.

| Material | $\rho$ ($10^3$ kgm$^{-3}$) | $C_p$ (Jkg$^{-1}$·K$^{-1}$) | $\kappa$ (Wm$^{-1}$·K$^{-1}$) | $\alpha$ ($10^5$ cm$^{-1}$) | r |
|---|---|---|---|---|---|
| Ta | 16.65 | 140 | 57 | 5.54 | 0.74 |
| Pt | 21.09 | 133 | 72 | 7.78 | 0.71 |
| CFB | 8.22 | 440 | 86.7 | 3.616[a] | 0.7[a] |
| Co | 8.90 | 421 | 100 | 7.54 | 0.72 |
| Fe | 7.874 | 449 | 80 | 1 | 0.65 |
| SiO$_2$ | 2.20 | 1052 | 1.4 | $10^{-11}$ | 0.045 |
| Si | 2.33 | 700 | 150 | $1.02\times10^{-2}$ | 0.33 |

$\rho$: density, $C_p$: specific heat capacity, $\kappa$: thermal conductivity, $\alpha$: absorption coefficient, r: reflectivity. [a]Estimated from compositional weight average of Co and Fe.

**Table II** | Numerically calculated temperature gradient and experimentally measured resistance.

|  | CFB/Pt | CFB/Ta | Pt/CFB | Ta/CFB | unit |
|---|---|---|---|---|---|
| $\Delta V R^{-1}$ | 8.920 | 5.096 | 3.568 | 1.510 | µVkΩ$^{-1}$ |
| $\nabla T_\text{F}$ | 7.840 | 5.824 | 2.181 | 2.134 | mKµm$^{-1}$ |
| $\nabla T_\text{FN}$ | -6.740 | -5.373 | 4.363 | 5.074 | mKµm$^{-1}$ |
| $R$ | 42.84 | 264.8 | 44.74 | 270.8 | Ω |
| $R_\text{N}$ | 46.26 | 487.8 | 48.49 | 508.5 | Ω |
| $\rho_\text{N}$ | 15.42 | 162.6 | 16.16 | 169.5 | µΩ·cm |